\documentclass[12pt]{iopart}

\usepackage{graphicx}

\newcommand{\lesssim}{\lower.5ex\hbox{$\; \buildrel < \over\sim \;$}}
\newcommand{\gtrsim}{\lower.5ex\hbox{$\; \buildrel > \over\sim \;$}}

\newcommand{\dD}{\delta_{\rm D}}
\newcommand{\g}{\gamma}
\newcommand{\apj}{{\it Astrophys.\ J.}}
\newcommand{\apjl}{{\it Astrophys.\ J.\ Lett.}}
\newcommand{\nat}{{\it Nature}}
\newcommand{\prd}{{\it Phys.\ Rev.\ D}}
\newcommand{\prl}{{\it Phys.\ Rev.\ Lett.}}
\newcommand{\ana}{{\it Astron.\ \& Astrophys.}}

\newcommand{\mnras}{{\it Monthly Not.\ Roy.\ Astron.\ Soc.}}

\begin{document}

\title[Ultra high energy cosmic rays from black hole jets of radio 
galaxies]{Ultra high energy cosmic rays from black hole jets of radio galaxies}

\author{C D Dermer$^1$, S Razzaque$^{1,2}$, J D Finke$^{1,2}$, 
A Atoyan$^{3}$}

\address{$^1$Code 7653, Naval Research Laboratory, Washington, DC 20375-5352 USA\\
$^2$Naval Research Laboratory/National Research Council Resident Research Associate\\
$^3$Concordia University, Montr\'eal, Quebec H3G 1M8 Canada
 }
\ead{charles.dermer@nrl.navy.mil}
\begin{abstract}
The Auger Collaboration reports \cite{aug07,aug08} 
that the arrival 
directions of $\gtrsim 60$ EeV ultra-high 
energy cosmic rays (UHECRs) cluster along the 
supergalactic plane and correlate with  
active galactic nuclei (AGN) within $\approx 100$ Mpc. 
The association of several events with 
the nearby radio galaxy Centaurus A supports 
the paradigm that UHECRs are powered by 
supermassive black-hole engines and 
accelerated to ultra-high energies in the 
shocks formed by variable plasma winds in the 
inner jets of radio galaxies. The GZK horizon length of 75 EeV 
UHECR protons is $\approx 100$ Mpc, so that the Auger results
are consistent with an assumed proton composition of the UHECRs.  
In this scenario, the sources of UHECRs are FR II radio galaxies
and FR I galaxies like Cen A with scattered radiation 
fields that enhance UHECR neutral-beam production. 
Radio galaxies with jets pointed away from us
can still be observed as UHECR sources
due to deflection of UHECRs by magnetic fields 
in the radio lobes of these galaxies. 
A broadband $\sim 1$ MeV  -- 10  EeV radiation component
in the spectra of blazar AGN 
is formed by UHECR-induced cascade radiation in the extragalactic 
background light (EBL). This emission is too faint to be seen from Cen A, but could be 
detected from more luminous blazars.
\end{abstract}

\submitto{\NJP}
\maketitle

\section{Introduction}

The Auger Observatory in the Mendoza Province of Argentina at 
$\approx 36^\circ$ S latitude
determines the arrival directions and energies of UHECRs 
using four telescope arrays to measure Ni air fluorescence 
and 1600 surface detectors spaced 1.5 km apart to measure muons 
formed in cosmic-ray induced showers. Event reconstruction using the hybrid technique 
gives arrival directions better than $1^\circ$, 
and energy uncertainties at $10^{20}$ eV (100 EeV) of $\sim 11$\% for a 
50\% Fe and 50\% p composition \cite{wat08}.  Analysis of the composition of the 
high energy showers in the early Auger analysis showed 
it becoming heavier, somewhere between p and Fe, at $10^{19.4}$ eV \cite{yam07,ung07}. 
By contrast, HiRes data are consistent with dominant
proton composition at these energies \cite{sok08}, but 
uncertainties in the shower properties \cite{ung08} and particle physics extrapolated 
to this extreme energy scale \cite{eng07} preclude definite statements 
about composition.

In the Auger analysis \cite{aug07,aug08}, a probability statistic $P$ corrected 
for exposure is constructed from the nearest-neighbor angular 
separation $\psi$  between the arrival direction of an UHECR 
with energy $E$ and the directions to AGN in the Veron-Cetty and 
Veron (VCV) catalog \cite{vcv06}, containing 694 active galaxies with $z <0.024$ or distance 
$d < 100$ Mpc. $P$ was minimized for  $\psi  = 3.1^\circ$, threshold clustering
energy
$E_{cl} = 56$ EeV, 
and clustering redshift 
$z_{cl} = 0.018$ ($d_{cl} \cong 75$ Mpc), containing 27 events
(the two highest energy events were 90 and 148 EeV).
Twelve events correlate within 3.1$^\circ$ of
the selected $d<75$ Mpc AGN, and another three within the vicinity
of one of these nearby AGN, ruling out an isotropic UHECR flux or a 
Galactic source population. Note that the VCV AGNs are not necessarily 
the sources of the UHECRs, but may only trace the 
same matter distribution as the actual UHECR sources.

This discovery opens the field of charged-particle
astronomy. Only at the highest energies can arrival directions
of charged particles be associated with their sources. This is because
deflections by (i) the Galactic 
or (ii) intergalactic magnetic (IGM) field isotropizes the directions
of lower energy cosmic rays. 

Here we consider the Auger clustering results within the paradigm that
extragalactic black-hole jet sources 
accelerate UHECRs. Estimates of deflection angle and
delays are given in Section 2, specialized to an assumed
proton composition of the UHECRs. The question of the GZK cutoff 
and the UHECR horizon is revisited in Section 3. 
In Section 4, we apply standard synchrotron theory to 
the lobes of Cen A in order to estimate the equipartition
magnetic field and absolute jet power, using a new technique 
that uses the 
jet/counter-jet ratio to determine the speed of the outflow. 
From this, limits on UHECR acceleration in colliding shells of 
blazars are used to derive maximum particle energies. Also, an estimate is
made of the flux of Compton-scattered 
cosmic microwave (blackbody) background radiation (CMBR), and compared with
the flux of secondary nuclear production in Cen A's radio lobes. 
In Section 5, we present calculations of the broadband 
$\gamma$-ray $\nu F_\nu$ flux from Cen A
due to secondary cascading of protons on the EBL, for $ 1$ nG
($= 10^{-9}$ G) IGM fields and show that Cen A is not detectable with current 
instrumentation.
Concluding remarks are given in Section 6.

\section{Magnetic Field Deflections of UHECRs}

For case (i), the Galactic magnetic field 
can be approximated by a magnetic disk with 
characteristic height $h_{md}$, giving a deflection angle 
$\theta_{dfl}\approx h_{md} \csc b/r_{\rm L}$. Here
$b$ is the Galactic latitude of the UHECR source, 
and the Larmor radius of a particle with energy $E$ 
and charge $Ze$ is 
\begin{equation}
r_{\rm L} \cong 65 \;{(E/60 {\rm EeV})\over ZB(\mu {\rm G})}\;{\rm kpc}\;.
\label{rL}
\end{equation}
 Thus 
\begin{equation}
\theta_{dfl,MW} \lesssim 1^\circ \; {Z h_{md}({\rm kpc})B(\mu {\rm G})
\over \sin b (E/60{\rm~EeV})}\;,
\label{thetadfl}
\end{equation}
limited by the finite extent of the magnetic disk.
Mean magnetic fields $B$ in the $\approx 0.2$ kpc 
thick gaseous disk of the Galaxy are $\approx 3$ -- 5 $\mu$G, 
but could fall to  $\ll 1 \mu$G 
in the kpc-scale halo \cite{aes02}. Deflection angles 
$\lesssim 3^\circ$ from Cen A ($b = 19.4^\circ$, 
galactic longitude $\ell = 309.5^\circ$, declination $-43^\circ$, 
distance $d \cong 3.5$ Mpc) restrict UHECRs to protons or 
light-Z nuclei and a small ($\lesssim 0.1~\mu$G) Galactic halo magnetic
field.

For case (ii),  the deflection angle of an UHECR ion 
when propagating through the IGM field
from a source at distance
$d$ is \cite{wax95,wc96}
$\theta_{d,IGM} \simeq d/ 2r_{\rm L}\sqrt{N_{inv} } 
\simeq {0.04^\circ }\;Z
\;\langle B_{-12} \rangle\;d(100 {\rm ~Mpc})/ [E(60{\rm~EeV})\sqrt{N_{inv}}]\;,
$
where $N_{inv}\simeq \max (d/\lambda,1)$ is the number of reversals  of
the magnetic field (also expressed through the magnetic-field correlation length
$\lambda$) and $\langle B_{-12} \rangle$  is the mean magnetic field of the 
IGM in pico-Gauss (1 pG = $10^{-12}$ G).  If UHECRs within 3$^\circ$ of Cen A are 
accelerated by the radio jets of Centaurus A, then to avoid
much larger deflections than those made by the 
Galactic magnetic field requires that 
$\langle B_{-12} \rangle \lesssim 2000\; \sqrt{N_{inv}}\;{E(60{\rm~EeV})/ [Zd(3.5 {\rm~Mpc})]}.$
By this reasoning, the mean IGM field in the directions towards AGNs 75 Mpc distant 
in the supergalactic plane (SGP) is restricted
to be $\lesssim 100\sqrt{N_{inv}}\;E(60{\rm~EeV})/ Z$ pG \cite{der07}. 
Time delays between electromagnetic outbursts and 
UHECR arrival windows from Cen A due to propagation through the IGM are about
\begin{equation}
\Delta t \approx {d^3\over 24 r_{\rm L}^2 c N_{inv}^{3/2}}\lesssim 
0.5\; {d^3(3.5 {\rm~Mpc}) Z^2 \langle B_{-12}\rangle^2 \over 
E^2(60 {\rm ~EeV})}\;{\rm~day}\;,
\label{Deltat}
\end{equation}
where the final expression holds because $N_{inv}\gtrsim 1$. 
Variable $\gamma$-ray flaring activity from Cen A could be reflected in variable
UHECR activity on sub-day timescales for IGM fields $\langle B_{-12} \rangle \lesssim 
10$  and $\lambda \lesssim 0.1$ Mpc or a large-scale ($\sim$ Mpc) 
ordered field with mean strength $\langle B_{-12} \rangle\lesssim 
1$.

The most intense magnetic fields  between us and Cen A 
consistent with the deflection data for an assumed proton composition 
of the UHECRs are 
$\langle B \rangle \approx 2 \; \sqrt{N_{inv}}$ nG. For the 
Auger data reaching to galaxies at distance $d \gtrsim 75$ Mpc, 
$\langle B \rangle \lesssim 100\; \sqrt{N_{inv}}$ pG.
For $N_{inv} \sim 100$ ($\lambda \cong 1$ Mpc), $\langle B \rangle \approx 1$ nG.
For the calculations in Section 5, we use $\langle B \rangle = 1$ nG which is a more
realistic value because UHECRs are correlated with more
distant sources as well.

\section{UHECR Proton Horizon}

The clustering energy $E_{cl}\cong 60$ EeV separates UHECRs formed mainly by sources
along the SGP at $d \lesssim d_{cl}$
from lower-energy UHECRs formed nearby and on 
$\gtrsim 75$ Mpc scales. 
A high-significance steepening in the UHECR spectrum at  $E\cong
10^{19.6}$ eV $\cong 4\times 10^{19}$ eV and
at $E\cong
10^{19.8}$ eV $\cong 6\times 10^{19}$ eV  was reported, respectively, 
by the Auger  \cite{yam07}
and  HiRes \cite{hires08} collaborations in 2007. 
These results confirm the prediction of Greisen, Zatsepin and Kuzmin 
\cite{gre66,zk66}
that interactions of UHECRs with CMBR photons cause a break in the UHECR 
spectral intensity near $10^{20}$ eV.

\begin{figure}[t]
\includegraphics[width=25pc]{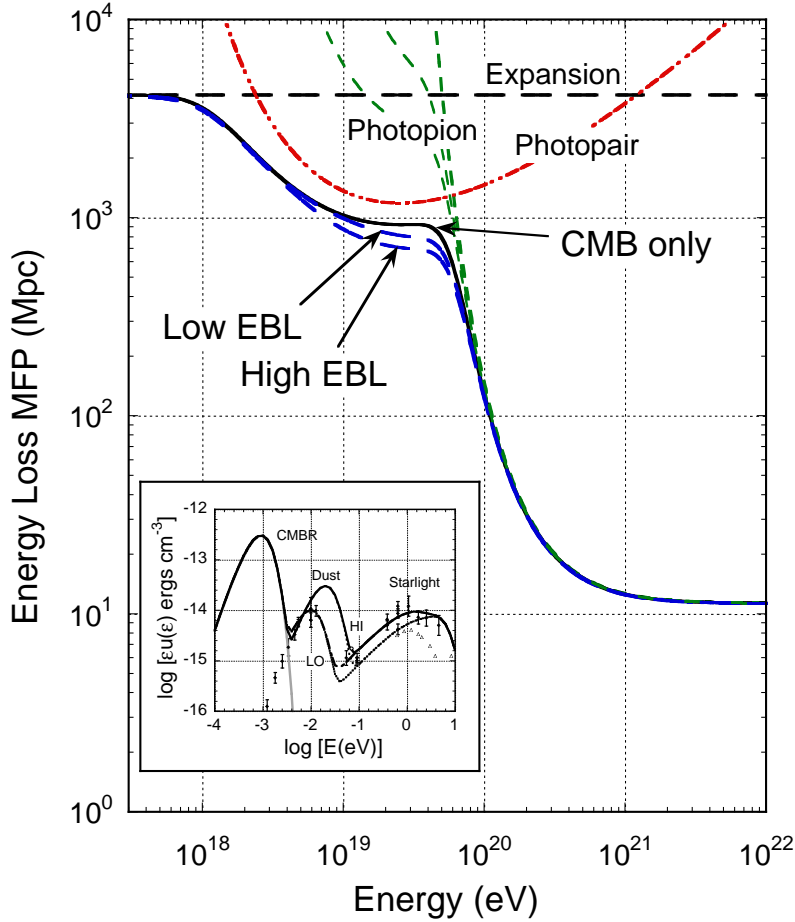}\hspace{-5pc}
\begin{minipage}[b]{16pc}
 \caption{\label{f1} 
Mean-free paths for energy loss of UHECR protons in different model
EBLs are shown by the solid curves, 
with photopair (dotted) and photopion (dashed)
components shown separately. ``CMB only" refers to total energy
losses with CMB photons only. 
Inset: Measurements of 
the EBL at optical and infrared frequencies, including phenomenological
fits to low-redshift EBL in terms of a superposition of modified blackbodies.
A Hubble constant of 72 km s$^{-1}$ Mpc$^{-1}$ is used throughout.\vskip1.0in 
}
\end{minipage}
\end{figure}

The clustering and GZK energies coincide because 
$\gtrsim 60$ EeV 
UHECR protons originating from sources 
at the $\gtrsim 100$ Mpc scale have lost a significant 
fraction of their energy due to photopion
losses with the CMBR, so that higher-energy particles
from the more distant universe cannot reach us. 
The energy-loss mean free
path $r_{\phi\pi} = c t_{\phi\pi} = c\gamma|d\gamma/dt |_{\phi\pi}^{-1}$
of an UHECR proton with energy $E=10^{20}E_{20}$ eV to photopion
losses with the CMBR at low redshifts 
is given, in good agreement with numerical calculations
\cite{ste68,sta00}, 
by the expression \begin{equation}
r_{\phi\pi}(E_{20}) \cong {13.7 \exp[4/E_{20}] \over [1 +
4/E_{20}]}\;{\rm Mpc}\;
\label{rphipi}
\end{equation}
\cite{der07}. The term $r_{\phi\pi}$ gives the mean distance over which a particle
with energy $E$ loses $\approx 1 - e^{-1}  \cong 63$\%
of its energy. The MFPs for energy loss by photopair and photopion losses in different
model EBLs, including an EBL consisting of the CMB alone, is shown in Fig.\ \ref{f1}.

\begin{figure}[h]
\center\hspace{-4pc}
\includegraphics[width=30pc]{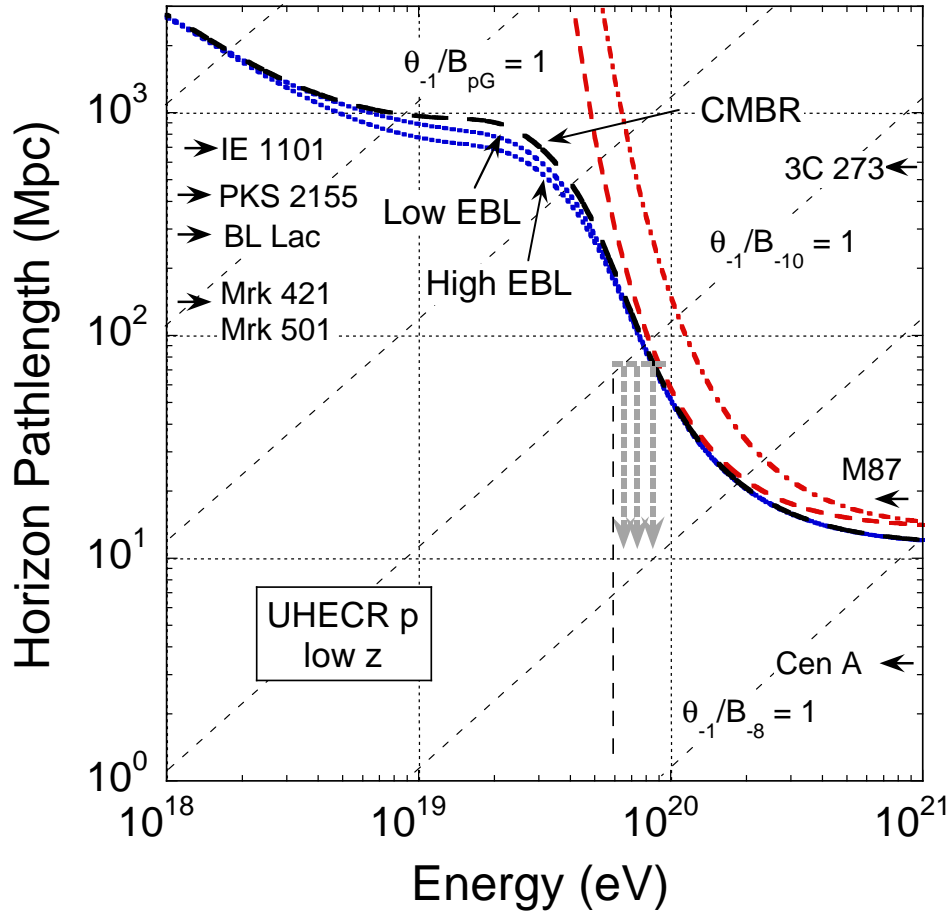}
\hspace{-8pc}
\caption{\label{f2} 
Heavy dotted curves give the 
horizon distance for UHECR protons as a function 
of total proton energy, using the local CMBR and
the low and high EBL target radiation field shown in the inset to Fig.\ \ref{f1}. 
Analytic approximation to photopion energy 
loss mean-free path, eq.\ (\ref{rphipi}), is
given by dot-dashed curve, and the UHECR proton horizon,
eq.\ (\ref{rhrz}), by
dashed curve, for CMBR only.  Auger clustering results are 
indicated by the shaded arrows. Distances to key radio galaxies 
are shown, and light dashed lines separate quasi-linear proton trajectories
from trajectories with strong deflections in overall ordered magnetic fields 
with strength
as labeled. Short-dashed lines give
length for a proton to be deflected by $0.1\theta_{-1}$ rad in magnetic
fields ranging from $10^{-8}$ -- $10^{-11}$ G.}
\end{figure}

The GZK horizon length giving the mean distance from 
which protons detected with energy $10^{20}E_{20}$ eV originate
depends in 
general on injection spectra
and source evolution \cite{hmr06} (see also \cite{ll08,ta08}), 
but a model-independent definition that
reduces to $r_{\phi\pi}(E_{20})$ for an energy-independent energy-loss rate
considers the average distance from which a proton with measured
energy $E$ had energy $eE$. The horizon distance, defined this 
way, is given by
$$r_{hrz}(E_{20}) \;=\; \int_{E_{20}}^{eE_{20}} {dx\over x} r_{\phi\pi}(x)\;\cong
$$
\begin{equation}
13.7\int_{E_{20}}^{eE_{20}}dx\;
{\exp(4/x)\over x(1+4/x)}\;{\rm Mpc}
\;\cong
{1.1E_{20}^2\exp(4/E_{20})\over 1+1.6 E_{20}^2/13.7}\;{\rm Mpc}
\;,
\label{rhrz}
\end{equation}
where the last two expressions give the proton horizon on CMBR photons alone.

Using the phenomenological fits to low and high EBLs at optical and IR frequencies,
represented in the inset to 
Fig.\ \ref{f1} as a superposition of blackbodies \cite{der07}, gives
the corrected horizon distance shown in Fig.\ \ref{f2}. 
The horizon distance for 
$\approx 57 (75)$ EeV UHECR protons is $\approx 200 (100)$ Mpc.
A GZK horizon smaller than 40 Mpc applies to protons
with $E \gtrsim 100$ EeV.  This explains the clustering observations
observed by the Auger collaboration if UHECRs are predominantly
protons, but would be inconsistent with the Auger results if the UHECRs are
composed of high-Z material, as the mean free path for photodisintegration can 
be much larger than 100 Mpc for $\approx 10^{20}$ eV ions like Fe \cite{der07}. If the Auger energy scale is
underestimated by $\approx 20$\% due to systematic effects (which 
would also reconcile the discrepancy between the different HiRes and 
Auger GZK energies 
as determined from the spectral break), then GZK losses on 
UHECRs with a dominant proton composition would explain
the clustering towards the SGP even more decisively. The higher energy
scale for the Auger experiment might also explain the lack of 
clustering observed by HiRes at $>57$ EeV \cite{abb08}, which would at this 
energy cut in the HiRes experiment include large numbers 
of lower-energy, more distant and less clustered cosmic rays.

\section{UHECRs from AGN Jets}

Discovery of UHECR arrival directions 
clustered towards the SGP was  
anticipated by  analysis of UHECR
 data from Haverah Park, AGASA, Volcano 
Ranch, and Yakutsk observatories  \cite{sta95}; see also \cite{ste68}. 
Compared to an isotropic source flux, 
the average and rms angular distances toward the SGP 
were enhanced at the 2.5 -- 2.8 $\sigma$ level for events 
with $E> 40$ EeV. Stanev et al.\ \cite{sta95} argued that their analysis
 favors radio galaxies as the sources of 
UHECRs. A radio-galaxy origin of UHECRs is 
consistent with harder radio sources 
\cite{sp89} and clustering of sources in the two Jansky,
 2.7 GHz Wall and Peacock catalog \cite{wp85} towards the SGP.

The SGP runs through the Virgo Cluster 
at $\approx 20$ Mpc, and contains an assortment 
of radio galaxies such as M87, Cen A and NGC 315, and the 
starburst galaxies M82 and NGC 253. Using infrared galaxy 
surveys to better define the SGP 
improves the significance of correlations of UHECRs with 
the SGP \cite{sta08}. When 
weighted by hard X-ray flux, the UHECR 
arrival directions are strongly correlated with 
Swift Burst Alert Telescope galaxies within 100 Mpc, which
trace the SGP \cite{geo08}. Searches \cite{mos08,zfg08}
for specific AGN in the VCV and NASA/IPAC NED
catalogs finds Seyfert 2, low ionization, and other radio-quiet 
galaxies closest to the UHECR arrival directions, 
in addition to associations with 
Cen A, Cen B, an FR II radio galaxy and a BL Lac object
within 140 Mpc. IGR J21247+5058, an 
FR II broad-lined radio galaxy at $z = 0.02$ or 
$d \approx 80$ Mpc, recently discovered with
INTEGRAL \cite{mol07}, is 2.1 degrees away from 
a HiRes Stereo event with $E > 56$ EeV
\cite{ad08}. At least 8 of the 27 UHECRs with $E>56$ EeV 
are within 3.5$^\circ$ of nearby radio galaxies \cite{nm08}.

We suppose that the evidence is compatible with an AGN origin 
of UHECRs in the black-hole jets of radio galaxies, with
Centaurus A being the most prominent example. Cen A itself, though
classified as an FR I radio galaxy, has a bolometric radio luminosity exceeding
$4\times 10^{41}$ ergs s$^{-1}$ \cite{alv00}, near the dividing
line between FR I and FR II galaxies in terms of radio power. The
giant elliptical galaxy in Cen A is intercepted by 
a small spiral galaxy making the prominent dust lanes \cite{eb83} which,
including inner dust torus emission,
could conceal the optical/UV line strengths and
distribution of broad line region gas 
near the Cen A nucleus \cite{gu01,zw06}. 
A strong scattered radiation field is important 
for photomeson losses and neutral beam
production in AGN jets \cite{ad01,ad03}, and seems required in 
many BL Lac objects because of spectral fitting
difficulties with synchrotron/SSC models \cite{gk03,gt08,fdb08}.

The nucleus of the Cen A jet is visible at radio and X-ray energies,
revealing sub-luminal ($v \sim c/2$) relativistic outflows and
jet/counterjet fluxes of X-ray knots consistent with mildly
relativistic speeds on projected $\sim 10$ pc scales \cite{kra02,har03}. Cen A was 
detected at hard X-ray and $\gamma$-ray energies with OSSE, COMPTEL 
and EGRET on the {\it Compton Observatory} \cite{kin95,sre99}, 
emitting $\approx 5\times 10^{42}$ 
ergs s$^{-1}$ in keV -- MeV radiation with day-scale variability observed
at $\approx 100$ keV (see also \cite{gup08}). Most of this emission is 
probably quasi-isotropic radiation from the hot accretion plasma.
Variability  of $\sim$ GeV radiation 
could be detected from Cen A with the {\it Fermi Gamma ray Space Telescope}
if the jetted $\gamma$-ray emission is not too dim. Such a 
detection would also discriminate 
between an inner jet and extended origin of the 
$\gamma$-ray emission, for example, 
from Cen A's lobes \cite{har08}. 

\subsection{Jet Power from Synchrotron Theory}

The corrected Auger point source
exposure is  $\chi\omega(\delta_s)/\Omega_{60} \cong 
(9000\times 0.64/\pi)$ km$^2$ yr, where $\chi$ is the exposure,
$\omega(\delta_s) \cong 0.64$
is an exposure correction factor for 
the declination of Cen A, and $\Omega_{60} \cong \pi$ is the 
Auger acceptance solid angle \cite{ch07}. For a power law proton injection spectrum
with number index $\alpha$,  the 
apparent isotropic UHECR luminosity from Cen A is
\begin{equation} 
 L_{{\rm Cen A}}\cong \;1.6\times 10^{6}\times 60^{\alpha - 1}
\;{4\pi d^2 N\Omega_{60}\over \chi \omega(\delta_s)}\;
\left({\alpha - 1\over \alpha - 2}\right)\;[E({\rm EeV})]^{2-\alpha}\;{\rm ergs~s}^{-1}\;,\;
\label{powercena}
\end{equation} 
which for $N = 2$ events 
gives $L_{{\rm Cen\,A}}(\alpha = 2.2)\simeq 5\times 10^{39} [E({\rm EeV})]^{-0.2}$ ergs s$^{-1}$
 and $L_{{\rm Cen\,A}}(\alpha = 2.7)\simeq 2\times 10^{40} 
[E({\rm EeV})]^{-0.7}$ ergs s$^{-1}$ \footnote{
This power should be increased
by factors 2 -- 3 if 5 or 6 events are associated with Cen A and its lobes \cite{mos08}.
}. 
The Cen A radio luminosity thus 
greatly exceeds the UHECR luminosity unless cosmic rays
are strongly beamed along the Cen A jet. 
The integrated cosmic ray power
for $\alpha =2.0$ from 1 GeV to $10^{20}$ eV is 
$\approx 2$ -- $3\times 10^{40}$ ergs s$^{-1}$ for $N = 5$ events. 

It is interesting to note that the production of 
$\approx 10^{40}$ ergs s$^{-1}$
in UHECRs by Cen A, the dominant radio galaxy 
within $\sim 10$ Mpc,
represents an UHECR 
emissivity of $\simeq 8\times 10^{43}$ ergs yr$^{-1}$ Mpc$^{-3}$ (cf.\ \cite{wb99}).

\subsubsection{Equipartition Magnetic Field}

The UHECR luminosity can be compared with the time-averaged
jet power inferred from the magnetic field energy in the radio lobes, 
assumed to be inflated by the pressure of the black-hole jet. 
Knowing the power of a black-hole jet allows 
one to derive the maximum particle energy of UHECRs accelerated
 through Fermi processes.

Here we state some elementary results from synchrotron theory
needed to derive the jet power \cite{dm09}. 
We treat a standard blob model and assume that the measured
radio emission is nonthermal synchrotron radiation from randomly 
oriented electrons in a randomly directed magnetic field. In this
approximation, the 
equipartition magnetic field is 
defined by equating the magnetic-field energy density $U_B = B^2/8\pi$
with the {\it total} particle energy density $U_{par}$
consisting of hadrons and electrons, 
assumed to be dominated by a large-scale, randomly
oriented magnetic field of strength $B$.

The equipartition magnetic field $B_{eq} = B(k_{eq} = 1)$ with
$U_{par} = k_{eq}U_B$, where
\begin{equation}
 B(k_{eq}) = 
{B_{cr}\over \dD}\;\left[
{3\pi d_L^2 m_ec^2 f^{syn}_{\epsilon_{2}}\ln (\epsilon_2/\epsilon_1)(1+\zeta_{pe})\over V_b^\prime  
c\sigma_{\rm T} U_{cr}^2 k_{eq} \sqrt{(1+z)\epsilon_{2}}}\right]^{2/7} 
\;.
\label{Beq}
\end{equation}
Here $B_{cr} = m_e^2c^3/e\hbar = 4.414\times 10^{13}$ G, 
$U_{cr} = B_{cr}^2/8\pi = 7.752\times 10^{25}$ ergs cm$^{-3}$,
and $\dD$ is the Doppler factor. In this expression, a factor $\zeta_{pe}$ more energy 
is assumed to be carried by protons and ions than leptons.
This result applies to the $F_\nu \propto \nu^{-1/2}$ portion of the
synchrotron spectrum made by electrons with $N^\prime(\gamma^\prime) 
\propto \gamma^{\prime -2}$ that carry most of the energy.

The $\nu F_\nu$ synchrotron flux $f_\epsilon^{syn}$ 
at photon frequency $\nu = m_ec^2 \epsilon/h$
is found in eq.\ (\ref{Beq}). 
Note that $f^{syn}_{\epsilon_{2}}/\sqrt{\epsilon_{2}}$
is constant throughout the range $\epsilon_1 \leq \epsilon < \epsilon_2$ when
$f^{syn}_{\epsilon}\propto \epsilon^{1/2}$, so that
 the value of $f^{syn}_{\epsilon }/\sqrt{\epsilon }$
is independent of 
$\epsilon$ over this range. Also, $B_{eq}$ depends only very
weakly on the normalization of the energy contained in electrons,
varying $\propto [\ln (\epsilon_2/\epsilon_1)]^{2/7}$.

\subsubsection{Synchrotron Power}

The total jet power of a one-sided jet, referred to the stationary black-hole reference
frame, is then
\begin{equation}
P^*_{j} \cong \pi r_b^{\prime 2} \beta c \Gamma^2 \;\left[{{B^2\over 8\pi} + {m_ec^2 (1+\zeta_{pe})\over 
V_b^\prime}}\;{6\pi d_L^2 f^{syn}_{\epsilon_{2}}\over 2 c\sigma_{\rm T} U_B \dD^4}\;
\sqrt{\dD\varepsilon_B\over \epsilon_{2}(1+z)}\ln({\epsilon_{2}\over \epsilon_1})\right]\;,
\label{Lj}
\end{equation}
defining $\varepsilon_B = B/B_{cr}$ 
From this, one can show \cite{dm09,fdb08} that
\begin{equation}
P_j^*(B) =  {3\over 7} P_j^*(B_{minL}) (u^2 + {4\over 3u^{3/2}})\;,
\label{LjstareB}
\end{equation}
and $\;u\equiv {B/B_{minL}}\;.$
The minimum power is 
\begin{equation}
P_j^*(B_{minL}) =  {7\over 3} \pi c \beta\Gamma^2 r_b^{\prime 2} U_{cr}
\left({B_{minL}\over B_{cr}}\right)^2\;.
\label{LjstareB1}
\end{equation}
The magnetic field giving the minimum jet power required to account for 
the synchrotron flux is 
\begin{equation}
B_{minL} = \big({3\over 4}\big)^{2/7}B_{eq} \cong 0.921 B_{eq}\;.
\label{BminL}
\end{equation}

\subsubsection{Application to Centaurus A}

The distance to Cen A is  
$d_{Cen~A} = 3.5 d_{3.5}$ Mpc, with $d_{3.5} \cong 1$. The preceding 
relations give
\begin{equation}
P_j^*(B_{minL}) =  {7\over 3} \pi c \beta\big( {\Gamma\over\dD}\big)^2 r_b^{\prime 2} U_{cr}
\left[{
27 d_L^2 m_ec^2 f^{syn}_{\epsilon_{2}}\ln (\epsilon_2/\epsilon_1)(1+\zeta_{pe})\over 16 
c\sigma_{\rm T} U_{cr}^2   \sqrt{(1+z)\epsilon_{2}}}\right]^{2/7}.\;
\label{LjstareB2}
\end{equation}
This applies to either spherical or cubical volumes by writing
the emitting volume in the fluid frame as 
\begin{equation}
V_b^\prime = g d_b^{\prime~3} = 6.7\times 10^{69}g (d_{3.5}\psi_0)^3
\;{\rm cm}^3\;.
\label{vbprime}
\end{equation}
The geometry factor is given by
$g = 1$ for cubical and $g = \pi/6$ for spherical volumes.
The angle $\psi_0$, in degrees, is the angular extent of the 
emitting volume.

Writing $f_\epsilon^{syn} = f_{\epsilon_2}\sqrt{\epsilon/\epsilon_2}$,
with $f_{\epsilon_2}=10^{-12}f_{-12}$ ergs cm$^{-2}$ s$^{-1}$, and 
normalizing to the 21 cm (1.43 GHz) flux,
so $\epsilon_2 \cong 1.15\times 10^{-11}\tilde\epsilon_{21}$,
then the magnetic field associated with the minimum jet power is
\begin{equation}
B_{minL} = {0.57\over \dD}\;
\left({f_{-12}\over d_{3.5}g\psi_0^3}\right)^{2/7}
\left( {\ln(\epsilon_2/\epsilon_1)(1+\zeta_{pe})\over
\sqrt{\tilde\epsilon_{21}(1+z)}}\right)^{2/7}\;\;\mu{\rm G}\;.
\label{BminLCenA}
\end{equation}
The minimum jet power for a one-sided jet is, therefore, 
$$P^*_j(B_{minL}) \cong 4\times 10^{43}\beta\left({\Gamma\over \dD}\right)^2\;
\times\;
$$
\begin{equation}
g^{2/21} d_{3.5}^{10/7}\psi_0^{2/7}\left( {\ln(\epsilon_2/\epsilon_1)(1+\zeta_{pe})\over
\sqrt{\tilde\epsilon_{21}(1+z)}}\right)^{4/7}\;{\rm ergs/s}\;.
\label{PminLCenA}
\end{equation}
Taking the bolometric factor $  
\ln(\epsilon_2/\epsilon_1)(1+\zeta_{pe}) = 10$
in eqs.\ (\ref{BminLCenA}) and (\ref{PminLCenA}) gives the results shown in Table 1
for the five different regions defined in Ref.\ \cite{har08}. 
Uncertainties by a factor $ 10$ in the bolometric factor translate to a factor 
$10^{2/7}\cong 1.93$ in $B_{minL}$ and a factor $10^{4/7}\cong 3.73$ in $P_{minL}$.

\Table{\label{tabl1}
Jet power of Cen A from synchrotron theory$^{\rm a}$}
\br
&&&\centre{2}{ }\\
Regions$^{\rm b}$ & F(Jy) & $f_{-12}$ & $\psi_0$ &g& $\beta$ & $B_{minL}$ & 
${P_j^*(B_{minL})\over (\Gamma/\dD)^2 (\lambda/10)^{4/7}}$\\
\ns
 & && & & & \crule{2}\\
\ns
& & & & & & $\mu$G & $10^{43}$ ergs s$^{-1}$ \\
\mr
1 &  $91.2\pm2.1$ & $1.3$ &  2.0 & 1 & 0.1 & 0.97 & \  2.1 \\

2 &  $74.6\pm3.0$ & $1.5$ &  1.0 & 1 & 0.2 & 0.99 & \ 3.4 \\

3 &  $545\pm1.3$ & $7.78$ &  2.0 & $\pi/6$ &  & $1.96/\Gamma$ &  \\

4 &  $204.7\pm4.3$ & $2.92$ &  1.7 & 1 & 0.2 & 1.26 & \ 6.2 \\

5 &  $111.2\pm6$ & $1.59$ &  1.6 & 1 & 0.1 & 1.08 & \ 2.2 \\
\mr
Total &  $1026.7\pm 8.3$ & $14.6$ &    &   &   &   &   \\

\br
\end{tabular}
\item[] $^{\rm a}$Radio observations at 21 cm from Ref.\ \cite{har08}.
\item[] $^{\rm b}$Regions defined in Ref.\ \cite{har08}.
\end{indented}
\end{table}

To get the speed of the outflow, we use the jet/counterjet 
ratio $\rho$, which is a number taken directly from observations.
If $\theta$ is the angle of the jet nearest to the line of sight, and 
we assume that all jets are two-sided with equal power ejected
in the opposite direction, then it is easy to show \cite{mr98,aa97}
that the speed 
\begin{equation}
\beta c = {c\over \cos\theta}\;\big({\rho^{2/7}-1\over \rho^{2/7}+1}\big)\;.
\label{betac}
\end{equation}
Thus we interpret the asymmetrical lobe flux in Cen A as a consequence of 
aberration of the mildly relativistic two-sided outflow, rather than as
a difference of spectral properties due to distinct environmental effects.
The jet power obtained from this interpretation can be compared
with other methods to determine jet power \cite{rs91}.

Table \ref{tabl1} shows that the jet/counterjet ratio $\rho \approx 1.5$ --
2 for regions 1 and 5, and $\rho \approx 2$ -- 4 for regions 2 and 4, 
each pair being at equal angular separation from Cen A's nucleus. 
The orientation $\theta$ of the jet of Cen A with respect to our line of sight
is $35^\circ \lesssim \theta \lesssim 72^\circ$ \cite{sdk94}, implying that $\beta \approx 0.1$
for regions 1 and 5, and $\beta \approx 0.2$ for regions 2 and 3, 
which are the values used in the calculations of the
jet power shown in Table 1. These speeds refer to the flow of 
plasma on the scale of hundreds of kpc from the nucleus of Cen A, 
so that the plasma from inner jet has to be ejected with much 
higher, probably relativistic speeds. The factor $(\Gamma/\dD)^2$, 
when included, reduces the power of the brighter lobe and increases the  
power of the dimmer lobe to reproduce the underlying assumption in the 
method that the powers of the two jets are equal. The minimum total jet power
to produce the weakly boosted Cen A radio emission is therefore $\approx 2\times 
4\times 10^{43}$ ergs s$^{-1}$, or total jet power $P_*^{tot}
\gtrsim 10^{44}$ ergs s$^{-1}$, as follows from Table 1.  

With total mean absolute Cen A jet powers $\approx 10^{44}$ ergs s$^{-1}$, 
apparent isotropic jet powers can reach  $10^{45}$ -- $10^{46}$ ergs s$^{-1}$ during
flaring intervals. Indeed, apparent isotropic
flaring luminosities in $\gamma$ rays alone exceed $10^{45}$ ergs s$^{-1}$
in Mrk 501 and Mrk 421, and $10^{46}$ ergs s$^{-1}$
in PKS 2155-304 \cite{aha07}. These AGN are nearby BL Lac objects that correspond to  FR I 
radio galaxies like Cen A seen along the jet axis, though Cen A may have the 
added advantage, in terms of UHECR production, of an external radiation field
to enhance photohadronic processes in the jet \cite{ad01,ad03}.
It will be interesting to apply this technique to a sample of radio galaxies,
including Perseus A (3C 84, NGC 1275; $z = 0.018$ or $d_L \cong 76$ Mpc) \cite{asa06,dkr98}, Cyg A \cite{ad08}, etc.

\subsection{UHECR Acceleration}

The particle energy 
density of a cold relativistic wind with apparent isotropic luminosity $L$ and Lorentz
factor $\Gamma = 1/\sqrt{1-\beta^2}$ at radius $R$
from the source is
$u_p = L/(4\pi R^2 \beta \Gamma^2 c)$. If a fraction $\epsilon_B$ is channeled
into magnetic field $B^\prime$ in the fluid frame, then $RB^\prime \Gamma = 
\sqrt{2\epsilon_B L/ \beta c}$, implying maximum particle energies 
$E^\prime_{max} \cong QB^\prime (R/\beta \Gamma)$, so
\begin{equation}
E_{max} \cong E^\prime_{max}\Gamma \cong \left({Ze\over \Gamma}\right) \sqrt{{2\epsilon_B L\over \beta c}}
\cong 2\times 10^{20} Z{ \sqrt{\epsilon_B L/10^{46}{\rm~ergs}{\rm~s}^{-1}}\over \beta^{3/2}\Gamma} \;{\rm eV}\;.\;
\label{Emax}
\end{equation} 
This simple, optimistic estimate begs the question how to transform directed
particle kinetic energy into magnetic field energy in a cold wind. In other words,
it is simply a dimensional analysis that contains no physical basis, whether for the 
source that emits a wind with such large apparent powers, or for the particle
acceleration mechanism.

Within the picture of particle acceleration through 
Fermi processes, a maximum particle energy
in colliding shells can be 
derived in a straightforward manner. 
The underlying
limitation of Fermi acceleration, whether first- or second-order, 
is that a particle cannot gain a significant fraction of its 
energy on a timescale shorter than the Larmor timescale. 
A colliding shell picture due to inhomogeneities in the relativistic
wind realistically applies to the inner jets of 
radio galaxies and blazars, and to GRBs.

Consider the ejection of two shells, 
with shell a ejected at stationary frame times 
$  0 \leq t_* < \Delta t_{*a}$, and  shell b at times 
$  t_{*d}\leq t_*  < t_{*d}+ \Delta t_{*b}$. The coasting Lorentz factor, 
wind luminosity (assumed constant during the duration of shell ejection),
and energy are $\Gamma_{a(b)}, L_{*,a(b)}$ and ${\cal E}_{*,a(b)}$, respectively, 
with stars referring to the stationary jet frame. 
For a collision, $\rho_\Gamma \equiv \Gamma_a/\Gamma_b < 1$. 
For relativistic winds, i.e., $\Gamma_ a\gg 1$, the
 collision radius is $r_{coll} \cong 2\Gamma_a^2 c (t_{*d} - \Delta t_{*a})$
when $\rho_\Gamma \ll 1$. When $\Gamma_b > \Gamma_a \gg 1$, the shocked
fluid Lorentz factor $\Gamma \gg 1$.

Collisions between relativistic shells divide into a number of cases
depending on whether the forward shock (FS) and reverse shock (RS) 
are relativistic or nonrelativistic \cite{der08}. We consider the case of a nonrelativistic
reverse shock (NRS) and relativistic forward shock (RRS), which is probably
most favorable for particle acceleration. Thus the FS Lorentz factor $\Gamma_f \gg 1$.
The magnetic field
of the forward-shocked fluid in the shocked fluid (primed) frame is
\begin{equation}
B^{\prime}_f = {\Gamma_f\over r_{coll}\Gamma_a}\;\sqrt{8\epsilon_{B,f}L_a\over c}\;,
\label{bprimef}
\end{equation}
and $\epsilon_{B,f}$, the $\epsilon_B$ parameter for the FS, is familiar from blast wave studies 
(e.g., \cite{spn98}).

The maximum particle energy in the comoving frame is $E_{max}^\prime \cong ZeB^\prime_f c \Delta t_a^\prime$, 
where $\Delta t_a^\prime$ is the comoving duration when particles are undergoing acceleration.
If this is equated with the time that it takes for the FS to pass through shell a, then
\begin{equation}
\Delta t^\prime_a \cong {\Gamma \Delta_a(r_{coll})\over \Gamma_f c}\;,
\label{deltatprimea}
\end{equation} 
From this, we obtain
$$E_{max} = \Gamma E^\prime_{max} \cong$$
\begin{equation}
 {Ze\over \left ( {t_{*d}\over \Delta t_{*a}}-1\right)}\;{\Gamma\over \Gamma_a^2}\;
\sqrt{{2\epsilon_{Bf}L_a\over c}}\lesssim {2.4\times 10^{20} Z }\;
{\Gamma\over \Gamma_a^2} \sqrt{\epsilon_{Bf}L_{46}}\;{\rm ~eV}.
\label{Emax1}
\end{equation}
The factor $\Gamma/\Gamma_a^2$ is at best of order unity, so this expression shows
that for protons to reach  $\gtrsim 10^{20}$ eV cosmic ray energies
 through Fermi processes, it is essential to consider sources with 
apparent isotropic luminosities  $\gtrsim 10^{46}$ ergs s$^{-1}$. 

Acceleration of UHECRs in Cen A is therefore in principle possible during
 powerful episodes of jet activity. Acceleration of UHECRs can also
take place in colliding winds of moderately
powerful blazars with $L\gtrsim 10^{46}$ ergs s$^{-1}$ 
where strong Doppler collimation takes place. 

\subsection{Deflection of UHECRs}

The milliarcsec-scale radio jets in Cen A show moderate
asymmetry, consistent with Cen A's radio jet being mildly relativistic 
and misaligned by $\sim 60^\circ$ \cite{hor06}. Off-axis radiation beaming factors
would conceal on-axis Cen A blazar-type flares with apparent 
powers $L\gtrsim 10^{46}$ ergs s$^{-1}$. These events
would eject cosmic
rays to $\gtrsim 10^{20}$ eV energies into the $\sim 100{\rm~kpc}\times 500$ kpc
radio lobe structure. If the equipartition field characterizes the
large-scale magnetic field of the lobes (i.e., $N_{inv}\sim 1$), then 60 EeV
UHECR protons with $r_{\rm L}\cong 65$ kpc (see eq.\ [\ref{rL}]),
 could be deflected
by the magnetic field of Cen A's lobe, or in the lobes of other
radio galaxies or BL Lac objects, as suggested 
in arrival direction maps \cite{mos08}.  

If UHECRs are formed through a neutron beam, then they will travel 
on average $\approx 500 [E/(60$ EeV)] kpc before decaying. Thus UHECRs will
be deposited throughout the radio lobe, with some not decaying until outside
the radio lobe structure.  Thus both radio galaxies and blazars can be 
sources of UHECRs. For distant sources, though, energy losses will reduce 
the arrival energies of UHECRs, so that searches for enhancements of UHECRs
towards specific sources should also take into account the EBL-dependent
horizon energy, as illustrated in Fig.\ \ref{f2}.

The proton Larmor radius is formally $r_{\rm L} \cong
110 E_{20}/B_{p{\rm G}}$ Gpc (eq.\ [\ref{rL}]), and photomeson interactions
with the CMBR and EBL take place on a horizon distance scale
$\approx r_{hor}(E_{20})$, where the received 
proton has lost $\approx 63$\%
of its energy. If the proton is hardly deflected out of 
the beam, so $d \ll r_{\rm L}$, then we should expect a pulse pile-up at a 
specific energy $r_{hor}(E_{20})\cong d_L$. For PKS 2155-304
at $z = 0.116$ or a propagation distance $d \cong 400$ Mpc (Fig.\ \ref{f2}), 
searches should be made
for enhancements of UHECRs with $E_{CR} \lesssim 40$ EeV, which is insensitive 
to the level of the EBL. 
If BL Lac objects accelerate UHECRs, then enhancements 
at $E_{CR} \sim 40$ -- 60 EeV can be searched for from Mrk 421 and Mrk 501.
Failure to detect a signal would constrain the minimum IGM field and number
of inversions, or call into question a radio/$\gamma$-ray galaxy origin of UHECRs.

\begin{figure}[h]
\hspace{8pc}
\includegraphics[width=30pc]{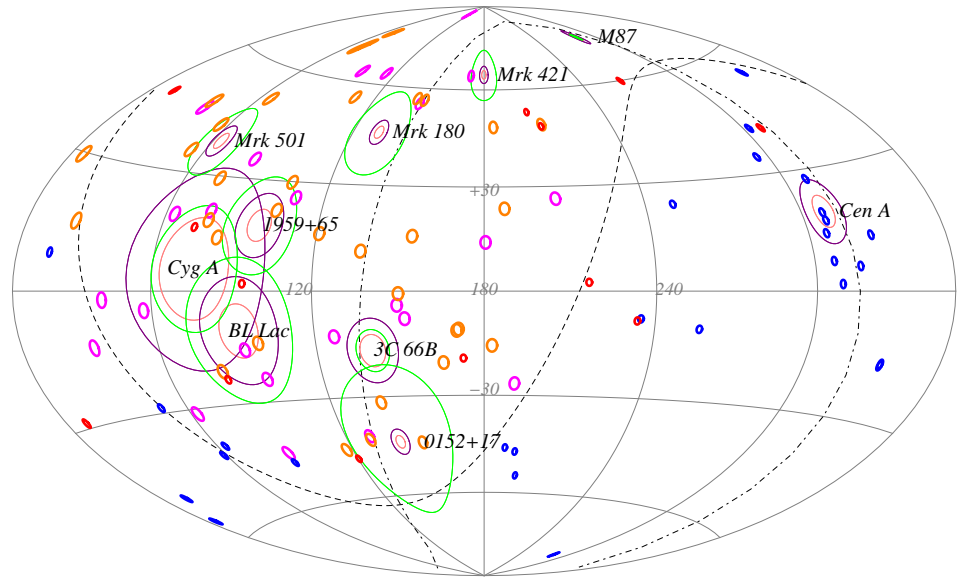}
\caption{\label{f3} Hammer-Aitoff projection in galactic coordinates of
UHECR arrival directions and directions to nearby prominent AGN; the 
direction to the Galactic Center 
is at the left and right extremities of this plot.  
Thick blue, red, and orange and magenta circles correspond to UHECRs
with $> 56$ EeV from Auger, $> 56$ EeV from HiRes, and $> 40$ EeV from AGASA, 
respectively.  The radii of the circles reflect the angular errors in 
the reconstructed cosmic ray directions, which is $1^\circ$ for both 
Auger and HiRes and $1.8^\circ$ for AGASA.   The centers of the concentric green, 
purple and pink circles correspond to the named AGN directions.  
The pink and purple circles represent angular deflections from the source AGN of the 
arriving cosmic rays with 40 EeV and 20 EeV energies, respectively, in the 
$\mu$G galactic disk magnetic field, using eq.\ (\ref{thetadfl}).  The green circles, 
on the other hand, represents angular deflection in an assumed $0.1$ nG intergalactic magnetic field, assuming
no magnetic-field reversals.  The dashed and dot-dashed curves correspond to the equatorial and super-galactic planes. }
\end{figure}

In Fig.\ \ref{f3} we plot the arrival directions of 27 cosmic rays from Auger (blue circles) and 13 cosmic rays  from HiRes (red circles), all with $> 56$ EeV energy and
with  $1^\circ$ angular uncertainty in their arrival directions (reflected in the radii of the circles; cf.\ Ref.\ \cite{vir02}).  We also plot the arrival directions of 58 AGASA UHECRs, consisting of 24 with $E > 56$ EeV (magenta circles)  and 
34 with $40$ EeV $< E < 56$ EeV (orange circles). Their arrival directions are displayed with an average angular resolution of 
$1.8^\circ$, as previously reported by the AGASA experiment \cite{hay99}. 
The number of AGASA UHECRs may suggest, given AGASA's exposure relative to HiRes and Auger, an energy calibration discrepancy between these experiments, which should be considered in more detailed studies but is ignored here.  Note that the arrival directions of the cosmic rays from the prominent nearby AGNs (labelled) are deflected by the galactic (pink and purple circles) and inter-galactic (green circles) magnetic field.  We used two energies, 40 EeV (pink circles) and 20 EeV (purple circles), to calculate deflections in the $\mu$G galactic magnetic field according to eq.\ (\ref{thetadfl}).  
The deflections from the AGN near the galactic plane (horizontal line) are larger because of the larger magnetic field.  

If Cyg A, BL Lac, 1959+65 or Mrk 501 are the possible sources of UHECR, then we should expect a large scattering, due to the galactic magnetic field,  in the arrival directions of
UHECRs below the GZK energy.   For distant sources, e.g., 0152+17, the deflection in the intergalactic magnetic field, plotted here for 0.1 nG field using green circles, may be larger.  However, reversals of the field orientation over $\sim$ Mpc scale may reduce the deflection in the intergalactic magnetic field from the values plotted in Fig.\ \ref{f3}.  These deflections in magnetic fields and related uncertainties can prevent positive identification of the UHECR sources below the GZK energy, although one should expect a clustering effect  as evident from the lower energy AGASA data in the northern hemisphere.

\subsection{ Nuclear $\gamma$ rays from Cen A's Radio Lobes}

Typical $\sim 1 ~\mu$G magnetic fields in Cen A's lobes 
carry an amount of energy $\cong V_{lobes} B^2/8\pi \cong
4\times 10^{57} V_{71} B_{\mu{\rm G}}^2$ ergs cm$^{-3}$, denoting
the Cen A lobe volume $ V_{lobes}=10^{71}V_{71}$ cm$^{-3}$.
For magnetic fields near equipartition, only $\approx $ few Myr
are required for a jet power of $\approx 10^{44}$ ergs s$^{-1}$
and total energy of $\approx 10^{58}$ ergs. Thus the Cen A
jet activity is likely to operate intermittently. 

Assuming at least equal energy in nonthermal protons and
ions as nonthermal electrons,
and using the equipartition assumption to normalize particle 
number and energy, we write the total cosmic ray 
spectrum over the Cen A lobes
as
\begin{equation}
N_{CR}(\g ) = {W_{CR}\over m_p({\rm ergs})}\;{p-2\over 1-\g_{max}^{2-p}}
\;\g^{-p}\;,\;{\rm for }\;1 \lesssim \g \lesssim \g_{max}
\;,
\label{NCR}
\end{equation}
with $\g_{max}\sim 10^{11}$ and total cosmic-ray energy $W_{CR} = 10^{58}W_{58}$ ergs.
A $\delta$-function approximation for the inelastic nuclear cross
section, $d\sigma_{pH\rightarrow \gamma}/dE_\gamma\cong
2\sigma_{\pi^0X}(\g )\delta (E_{\gamma} - \xi E_p),$
where $\sigma_{\pi^0X}(\g ) \cong 27\ln \g + 58/\sqrt{\g } - 41 $ mb is the
$\pi^0$ inclusive cross section \cite{der86} and $\gamma$-ray 
secondary fractional energy 
$\xi\sim $ 5 -- 10\%, gives a lobe-integrated 
$\gtrsim 1$ GeV $\nu F_\nu$ spectrum
\begin{equation}
f_\epsilon^{pp\rightarrow \gamma}
= \nu F^{pp\rightarrow \gamma}_\nu\;
= \;{\xi c n_{\ell}\over 2\pi d^2}\;W_{CR}(p-2) \sigma_{\pi^0}(\tilde \gamma)\;,\;
\tilde \gamma \equiv \left({E_\gamma\over \xi m_pc^2}\right)\;,
\label{xyz}
\end{equation}
so
$$f_\epsilon^{pp\rightarrow \gamma}\;({\rm ergs~cm}^{-2} {\rm ~s}^{-1})\;\simeq  $$
\begin{equation}
{4\times 10^{-12}\xi^{p-1}(p-2)\over 0.94^{2-p}}\;\left({n_\ell\over 10^{-4}~{\rm cm^{-3}}}\right)\;{ W_{58}\over d_{3.5}^2}\;
\left[ {\sigma_{\pi^0X}(E_\gamma/\xi m_pc^2 )\over 100~{\rm mb}}\right]
\;E_\gamma^{2-p}\;.
\label{nuFnupp}
\end{equation}
For $\xi = 0.05$ and $p \cong 2.3$, 
the nuclear $\gamma$-ray flux from the lobes of Cen A is
\begin{equation}
f_\epsilon^{pp\rightarrow \gamma}\;\simeq {2.4\times 10^{-14}W_{58}E_\gamma^{-0.3}\over 
\;d_{3.5}^2 }\;\left({n_\ell\over 10^{-4}~{\rm cm^{-3}}}\right)\;
\left[ {\sigma_{\pi^0X}(20E_\gamma )\over 100~{\rm mb}}\right]
\;{{\rm ergs} \over {\rm cm}^{2} {\rm ~s}}\;,
\label{nuFnupp2}
\end{equation}
with $E_\gamma$ now in GeV. Based on lack of observed internal depolarization
and measured soft X-ray flux, Ref.\ \cite{har08} 
argue that thermal particle target densities 
$n_\ell\sim 10^{-4}$ cm$^{-3}$.

The number of source counts with $E_\g$(GeV) $ \geq E_1$ detected with the
Fermi Telescope for $\xi = 0.05$ and $p \cong 2.3$ is
$S(>E_1 ) \cong X\Delta t f_{GeV}\int_{E_1}^\infty dE_\g E_\g^{a-2} A_{G}(E_\g) $
$\approx 0.7(X/0.2)\Delta t({\rm yr}) (n_\ell/10^{-4} {\rm~cm}^{-3}) W_{58}E_\g^{-1.3}$, where the 
effective area of the Fermi Telescope is
$A_G(E_\g ) \cong 8500/\sqrt{E_\g}$ cm$^{2}$ below 1 GeV and $A_G(E_\g ) \cong 8500$
cm$^{2}$ at higher energies, and $X\approx 1/5$ in the scanning mode. 
The number of background counts from the angular extent, 
$\Delta \Omega \cong 12/57.3^2$, of the lobes of Cen A, using as background
the diffuse extragalactic $\gamma$-ray background of 
Galactic background \cite{sre98}, is $B(>E_\g )
\cong 280 (X/0.2)\Delta t({\rm yr}) E_\g^{-1.1}$\footnote{The Galactic diffuse $\gamma$-ray background has a softer spectrum
than the extragalactic background, 
so is probably not significant at $E_\g\gg 1$ GeV
at the $\approx +20^\circ$ galactic latitude of Cen A.} 
.
Unless $n_{\ell}\gg 10^{-4}$ cm$^{-3}$, this emission signature would 
be too faint to be detected with 
the Fermi Gamma ray Space Telescope.  Thus the cosmic-ray induced nuclear emission 
in the lobes of Cen A is probably entirely negligible. The much more optimistic 
estimates in Ref.\ \cite{kot08} are due to the assumption of a very soft injection 
index $p \cong 2.7$ extending from GeV to ZeV energies, which requires 
a much larger cosmic-ray power.


\subsection{Thomson-Scattered CMBR}

The same electrons that make the radio synchrotron radiation also upscatter 
photons of the surrounding 
radiation fields. Using a $\delta$-function approximation for Thomson scattering 
of quasi-monochromatic photons with mean energy $m_ec^2 \epsilon_o$  and energy density $U_o$ gives the Thomson-scattered
$\nu F_\nu$ flux
\begin{equation}
f_\epsilon^{\rm T} \cong \dD^2 \left( {U_{o}\over U_B}\right )
f^{syn}_{\epsilon_{pk}}\sqrt{y} H(y;0,1)\;,\;y = {\epsilon\varepsilon_B\over 2\dD\epsilon_{o}\epsilon_{pk}}\;;
\label{feT}
\end{equation}
the value of the Heaviside function $H(y;0,1)= 1$ when $0\leq y< 1$ and 
$H(y;0,1)= 0$
otherwise.
For the CMBR, $m_ec^2 \epsilon_o = 1.24\times 10^{-9}
m_ec^2 = 0.63$ meV $ = 153$ GHz and $U_0 \cong 4\times 10^{-13}(1+z)^4 $ ergs cm$^{-3}$
(recall that $\varepsilon_B = B/B_{cr}$). Thus
\begin{equation}
f_\epsilon^{\rm T} \cong {10\dD^2\over B_{\mu\rm{\rm G}}^2}
f^{syn}_{\epsilon_{pk}}\sqrt{{\epsilon\over \epsilon_{\rm T}}}\;H(\epsilon;0,\epsilon_{\rm T})\;.
\label{feT1}
\end{equation}
For the CMBR, the mean dimensionless energy of a photon that is scattered 
by an electron radiating most of its synchrotron emission at $\epsilon_{pk}$
is
\begin{equation}
\epsilon_{\rm T}= {2\dD \epsilon_o \epsilon_{pk}\over \varepsilon_B}\;,
\label{epsT3}
\end{equation}
implying emission at
\begin{equation}
\;E_{\rm T}\cong  43 {\dD\over  B_{\mu\rm{\rm G}}}\left({\nu_{pk}\over 100 {\rm ~GHz}}\right)\;{\rm MeV}\;.
\label{epsT4}
\end{equation}
The WMAP observations at  22.5, 32.7, 40.4, 60.1, and 92.9  GHz 
  used by Hardcastle et al.\ \cite{har08} allowed these authors to predict 
the Thomson-scattered flux. As can be seen from this approximation,
this flux extends only to the low-energy end ($\lesssim 100$ MeV) 
of the Large Area Telescope frequency
range on the Fermi Gamma ray Space Telescope \cite{che07}.

\section{Cascade $\gamma$ rays from UHE Neutral Beams}

For flares in BL Lacs exceeding $\gtrsim 10^{46}$ 
ergs s$^{-1}$, and the much more energetic flares in FR II galaxies 
and flat-spectrum radio quasars reaching
apparent $\gamma$-ray powers $\gtrsim 10^{48}$ ergs s$^{-1}$, we describe
the sequence of events starting to produce cascade radation
following the acceleration 
of particles in black-hole jets \cite{ad03}. A neutral beam is formed
as a consequence of photomeson interactions of the accelearted UHECR protons 
with the synchrotron and scattered radiation
field in the inner jet. Some $\gtrsim 10$\% of the hadronic energy can 
be transformed into an escaping neutron beam, with a few percent
in a UHE $\gamma$-ray beam and a few percent into neutrinos \cite{ad01,ad03}.
Escaping neutrons with energy $E_n$ travel 
$E_n/10^{20}$ eV Mpc before decaying into protons and 
low-energy $\beta$-decay leptons and neutrinos. 

\begin{figure}[h]
\includegraphics[width=25pc]{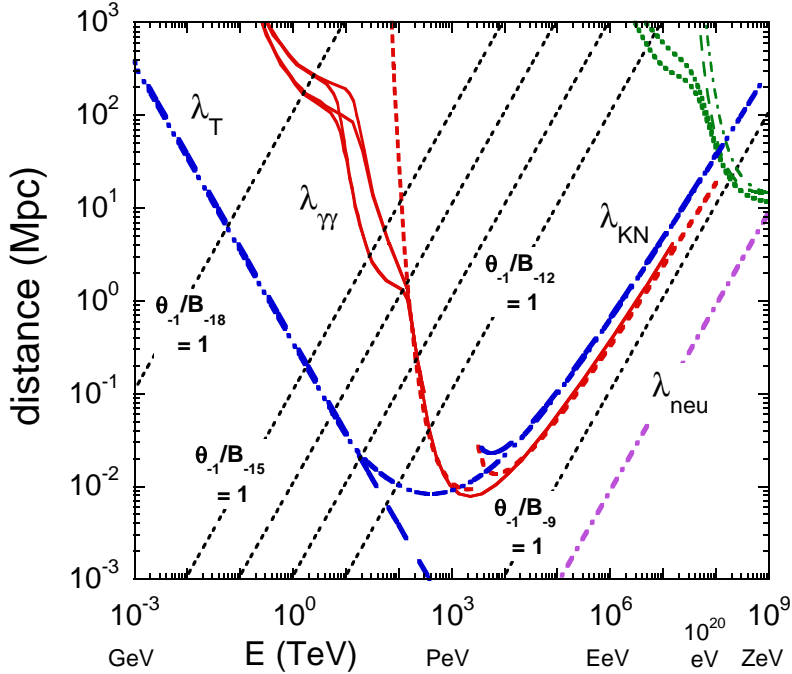}\hspace{-5pc}
\begin{minipage}[b]{17pc}\caption{\label{f4}
Mean-free paths
for attenuation of $\gamma$ rays by the EBL ($\lambda_{\g\g}$),
  for Compton-scattering
energy loss of relativistic electrons or positrons
with the CMBR in the Thomson ($\lambda_{\rm T}$)
and Klein-Nishina ($\lambda_{\rm KN}$)
regimes, and neutron decay ($\lambda_{neu}$). 
The photopion energy-loss pathlength for protons,
from Fig.\ \ref{f1}, are shown for comparison. The 
particle or photon energy is denoted $E$(TeV). Short-dashed lines give
length for a lepton to be deflected by $0.1\theta_{-1}$ rad in a magnetic
field from $10^{-9}$ -- $10^{-18}$ G, as labeled. Neutron decay length
is given by $\lambda_{neu}$.
}
\end{minipage}
\end{figure}

The $\g$ rays formed by $p,n+\gamma\rightarrow
\pi^0\rightarrow 2\g$ processes in the inner jet and 
particle beam are attenuated by CMBR photons
through  $\g\g \rightarrow e^{\pm}$ on length scales
$\lambda_{\g\g}^{bb}($kpc)$\cong 2 E_{PeV}/\ln(0.4 E_{PeV})$ kpc
for $E_{PeV} \gg 1$ and
$\lambda_{\g\g}^{bb}($kpc)$\cong 4\sqrt{E_{PeV}}\exp(1/E_{PeV})$ kpc
for $E_{PeV} \ll 1$, valid until $E_{PeV}<0.1$, when absorption
 on the extragalactic background light from stars and
dust dominates. Leptons with $\g \equiv 10^9\g_9$ Compton-scatter
the CMBR photons, losing energy on length scales
$\lambda_{\rm T} \cong 0.75/\g_9$ kpc when $\g_9 \ll 1$, 
and $\lambda_{\rm KN} \cong 2.1\g_9/[\ln(1.8\g_9)-2]$
 kpc when $\g_9 \gg 10$.

If the distance $d$ to the source is smaller than 
the particle Larmor radius and the 
correlation length (formally, $N_{inv} = 1$), 
the cascade pairs are deflected out of a beam with opening
angle $\theta = 0.1\theta_{-1}$ radian when $\theta r_{\rm L} < 
\lambda_{\rm T}$, implying that the collimation of the electromagnetic
cascade is preserved until $\g_9 \cong 0.4\sqrt{B_{-11}/\theta_{-1}}$.
Before the beam disperses, 
CMBR photons are scattered to energies $\epsilon \sim 10^{-9}\g^2$,
or $E_\g \cong 100 B_{-11}/\theta_{-1}$ TeV. 
Fig.\ \ref{f4} shows that the ratio $B_{-11}/\theta_{-1}\lesssim 0.1$ is 
unique in that Compton-scattered CMBR
from the cascade emerges from behind the EBL absorbing screen
to form a hard $\gamma$-ray component at 10 GeV -- 10 TeV 
energies. If $B_{IGM} \lesssim 10^{-12}$ G, then the cascade emission
from sources like 1ES 1101-232, if UHECR sources, could make
an anomalous hard $\gamma$-ray component. 

\begin{figure}[t]
\includegraphics[width=26pc]{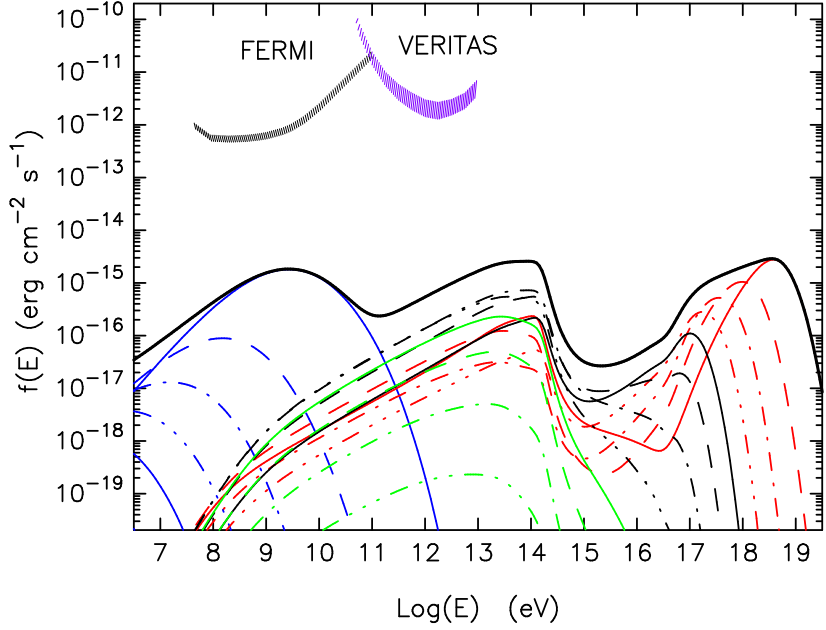}\hspace{-5pc}%
\begin{minipage}[b]{15pc}\caption{\label{f5}
Rectilinear cascade formed by photohadronic processes from UHECRs
produced by Cen A. Lower-energy blue family of curves shows the synchrotron
emission from successive generations of the cascade, and higher-energy 
family shows various generations of the Compton/$\gamma$-$\gamma$ cascade.
One-year Fermi Gamma ray Space Telescope/GLAST and 50 hr 
VERITAS sensitivities are shown for comparison.
}
\end{minipage}
\end{figure}

The high-energy cascade flux induced by rectilinear motions of
UHECRs from Cen A towards us is shown in Fig.\ \ref{f5}. 
This calculation is  a one-zone model, because it assumes
that the properties of the IGM remain unchanged between Cen A 
and the Galaxy. Cosmic ray propagation through the extended magnetized environment 
in the cocoon and radio lobes of Cyg A \cite{ad08} can also produce a $\gamma$-ray 
signature, but Cyg A is beyond the GZK horizon, so only lower energy 
UHECRs can reach us from this source. (For UHECRs accelerated in galaxy
clusters \cite{ias05} or other accelerators see, e.g.,  \cite{ga05}.)
This calculation is normalized to 2 protons   from Cen A with  $E >60$ EeV
for the exposure and area of the Auger observatory (eq.\ [\ref{powercena}]).
  
The parameters of this calculation are 
$d=3.5$ Mpc, $B=10^{-9}$ G, and particle injection
index $ \alpha = 2.2$. 
The proton injection spectrum is exponentially
cutoff at energy  $E_{max} = 2\times 10^{20}$ eV.
The EBL used is from Ref.\ \cite{pri05}. 
Twelve cascade cycles are sufficient to 
reach convergence, and this is the number used to calculate the total
Cen A spectral energy distribution induced by UHECRs interacting
with intervening radiation fields. The synchrotron cascade radiation
is colored blue. For the assumed magnetic field, the synchrotron flux is produced by  $\gamma \gtrsim 10^{12}$ 
    electrons and remains collimated. However, the Compton flux between $\approx 1$ - 100 TeV 
    does not take into account the deflection of electrons. It should therefore be considered
     as an upper limit that could be reached if $B \lesssim 10^{-12}$ G.

The solid curve in Fig.\ \ref{f5} shows the multiwavelength $\nu F_\nu$ flux,
but is clearly far too faint to be seen with present technology.
In this case, the proximity of Cen A works against it being a 
bright hadronic source: there is not enough pathlength to extract 
a significant fraction of the UHECR proton's energy. 
In future work (Atoyan et al.\ 2009, in preparation),
details of cascades formed by UHE leptons and $\gamma$-rays 
formed by photohadronic processes will be presented. The prospects of 
detecting more distant, aligned blazar jets, if they are sources
of UHECRs, could be more favorable if $B\lesssim 10^{-12}$ G.
UHECR production
in luminous  flat spectrum radio quasars 
could make anomalous hard
components at multi-GeV energies to be detected 
with the Fermi Gamma ray Space Telescope. Such 
emission signatures could also be seen in GRBs
\cite{dl02,rmz04}. 

\section{Summary and Conclusions}

We have considered the implications of the 
discovery \cite{aug07,aug08} by the Pierre Auger collaboration
of clustering of UHECR arrival
directions towards Cen A and AGN in the SGP. 
To guide out thinking, we have introduced a model-independent
definition of the GZK horizon distance, shown 
in Fig.\ \ref{f2} (cf.\ \cite{hmr06,ll08}). Assuming that 
black-hole jets energize and inflate the radio lobes of 
radio galaxies, a new method to calculate jet power
based on the radio spectrum, size, and jet/counterjet ratio is applied
to Cen A data. The average absolute jet power of Cen A is found
from this technique to be $\approx 10^{44}$ ergs s$^{-1}$. The apparent
isotropic power in a small beaming cone could easily, therefore, exceed
$10^{46}$ ergs s$^{-1}$ during flaring intervals.

An apparent jet power at least this great is needed to accelerate UHECRs
through Fermi processes. 
The collimated UHECRs, consisting mainly of neutron-decay protons,
 can be deflected by the $\approx \mu$G fields 
in the lobes of radio galaxies (for acceleration in the lobes of radio 
galaxies see, e.g., \cite{fm08}). This can give the appearance that UHECRs are 
emitted from Cen A's radio lobes \cite{mos08}, or allow UHECRs to originate
from more distant, misaligned radio galaxies like Cyg A, which itself has
a highly magnetized, $\approx 20~\mu$G cavity within several hundred kpc 
of the central engine \cite{ad08}. The enhancement of UHECRs
in the directions to specific radio jet sources at different redshifts
would give valuable information about the IGM magnetic fields and 
intensity of the EBL. 

The inner jets of radio galaxies, including Cen A, 
can make an escaping neutral beam of UHECRs. 
Because Cen A is a nearby radio galaxy with its closer jet misaligned
by $\sim 35^\circ$ -- $70^\circ$ to the line of sight, the UHECR
flux received from it is small, and the cascade radiation flux
is not detectable with current instrumentation. Neither is the 
secondary nuclear $\gamma$-ray emission from cosmic ray interactions
with the lobe's thermal gas if its density is $\approx 10^{-4}$ cm$^{-3}$
\cite{har08}, though the pair halo flux could be detectable \cite{acv94,sta06}.
The most prominent multi-MeV radiation
signature is due to the CMB photons Compton-scattered by the radio-emitting
electrons to soft $\gamma$-ray energies \cite{che07,har08}. 
Use of the transient event class in analysis of Cen A data \cite{atw08} could
help pull out the low-energy, $\approx 10$ -- $100$ MeV signal, which would
be important for
normalizing its magnetic field and total jet power. 
 
For UHECR blazar sources pointed towards us, 
anomalous hard cascade $\gamma$ radiation spectra
is potentially detectable with the Fermi Gamma ray Space Telescope 
or ground-based $\gamma$-ray telescopes, as we show in detail in future work. 
Detection of anomalous $\gamma$-ray signatures in blazars 
(or GRBs) could reveal 
emissions from UHECR acceleration in these sources. 
Lack of association of UHECR arrival directions with radio galaxies could
require that various other classes of extragalactic bursting sources be 
admitted, including 
long duration GRBs \cite{wda04,mur08}, 
low luminosity GRBs \cite{wan07}, or magnetars. 
For instance, Ghisellini et al.\ 
\cite{ghi08} argue that the Auger UHECRs are not actually associated with
Cen A, but with the background Centaurus cluster at $\approx 40$ -- 50 Mpc.
They argue that the rapid discharge of young, highly magnetized pulsars, 
which are the progenitors of magnetars, might have sufficient power to explain UHECR origin. 
Here the future pattern of UHECR arrival directions found with the 
Pierre Auger Observatory is of great interest,
and whether a concentration builds toward the direction to Cen A.

Spectral signatures associated 
with UHECR hadron acceleration in
studies of radio galaxies and blazars 
 with the Fermi Gamma ray Space Telescope and 
ground-based $\gamma$-ray observatories can provide evidence 
for cosmic-ray particle acceleration in black-hole plasma jets. Together with
IceCube or a northern hemisphere neutrino
telescope, observations of PeV neutrino  and MeV -- GeV -- TeV $\gamma$ rays
can confirm whether black-hole jets in radio galaxies 
accelerate the UHECRs.

\vskip0.2in
\noindent 
Useful discussions and correspondence
with Roger Blandford, Teddy Cheung, Diego Harari, 
Vasiliki Pavlidou, Esteban Roulet, and Alan Watson
are gratefully acknowledged. We thank Dr. Floyd Stecker 
for alerting us to an error in the calculations of the photopion losses of protons in the EBL.
 Visits of AA 
to the Naval Research Laboratory and research of  JDF are supported
by NASA  GLAST Interdisciplinary Science Investigation Grant DPR-S-1563-Y. 
The research of CD is supported by the Office of Naval Research.

\section*{References}


\end{document}